\documentclass{revtex4-1}
\usepackage[latin9]{inputenc}
\usepackage{color,graphicx,epsfig}
\usepackage{ifpdf}
\usepackage{amsmath}
\usepackage{bm}
\usepackage{color}
\usepackage[english]{babel}
\usepackage{graphicx}%
\usepackage{amsfonts}%
\usepackage{amssymb}
\usepackage{braket}
\usepackage{hyperref}

\definecolor{nicered}{rgb}{0.7,0.1,0.1}
\definecolor{nicegreen}{rgb}{0.1,0.5,0.1}
\hypersetup{colorlinks,citecolor= nicegreen,linkcolor= nicered}

\makeatletter

\DeclareFontEncoding{LGR}{}{}
\DeclareTextSymbol{\~}{LGR}{126}
\newcommand{\lyxmathsym}[1]{\ifmmode\begingroup\def\b@ld{bold}
  \text{\ifx\math@version\b@ld\bfseries\fi#1}\endgroup\else#1\fi}

\@ifundefined{textcolor}{}
{%
 \definecolor{BLACK}{gray}{0}
 \definecolor{WHITE}{gray}{1}
 \definecolor{RED}{rgb}{1,0,0}
 \definecolor{GREEN}{rgb}{0,1,0}
 \definecolor{BLUE}{rgb}{0,0,1}
 \definecolor{CYAN}{cmyk}{1,0,0,0}
 \definecolor{MAGENTA}{cmyk}{0,1,0,0}
 \definecolor{YELLOW}{cmyk}{0,0,1,0}
}

\makeatother

\begin{document}


\title{
 Isospin violating decays of positive parity $B_s$ mesons in HM$\chi$PT}
\author{ Svjetlana Fajfer}
\email[Electronic address: ]{ svjetlana.fajfer@ijs.si}
\affiliation{Department of Physics,
  University of Ljubljana, Jadranska 19, 1000 Ljubljana, Slovenia}
\affiliation{J. Stefan Institute, Jamova 39, P. O. Box 3000, 1001
  Ljubljana, Slovenia}

\author{ A. Prapotnik Brdnik}
\email{ anita.prapotnik@um.si}
\affiliation{Faculty of Civil Engineering, Transportation Engineering and Architecture, University of Maribor, Smetanova ulica 17,
2000 Maribor, Slovenia}

\begin{abstract}
Recent lattice  QCD results  suggest that the masses of the first two positive parity $B_s$ mesons lie below the $BK$ threshold, similar to the case of $D^*_{s0}(2317)^+$ and $D_{s1}(2460)^+$ mesons. The  mass spectrum of $B_s$ mesons  seems to follow pattern of  $D_s$ mass spectrum.  As in the case of charmed mesons, the  structure of positive parity $B_s$ mesons is very intriguing. To shed more light on this issue, we investigate  strong  isospin violating decays  $B_s(0^+) \to B_s^0 \pi^0$, $B_s(1^+) \to B_s^{*0} \pi^0$ and $B_s(1^+) \to B_s^0 \pi \pi$ within  heavy meson chiral perturbation theory. 
The two body decay amplitude arises at  the tree level and we show that the loop corrections give significant contributions. On the other hand, in the case of three body decay 
$B_s(1^+) \to B_s^0 \pi \pi$ amplitude occurs only at the loop level.   We find that the decay widths for these decays are: $\Gamma (B_s(1^+) \to B_s^0 \pi \pi)\sim  10^{-3}\,$keV and $\Gamma (B_s(0^+) \to B_s^0 \pi^0) \leq 55\,$keV, $\Gamma (B_s(1^+) \to B_s^{*0} \pi^0) \leq 50\,$keV. 
More precise knowledge of the coupling constant describing the interaction of positive and negative parity heavy  mesons with light pseudoscalar mesons would help to increase accuracy of our calculation. 
\end{abstract} 
\maketitle

\section{Introduction}

Two positive parity mesons $B_s$ states: the $J^P=1^+$ state $B_{s1}(5830)^0$ and the $J^P=2^+$ state $B^*_{s2}(5840)^0$ were observed by the CDF and LHCb collaborations
 \cite{Abazov:2007af,Aaltonen:2007ah,Aaij:2012uva,Aaltonen:2013atp}. 
 Recent lattice results \cite{Lang:2015hza}, as well as  previous works  of the authors 
\cite{Colangelo:2012xi,Kolomeitsev:2003ac, Cleven:2010aw, Altenbuchinger:2013vwa, Gregory:2010gm, Guo:2006fu, Guo:2006rp, Bardeen:2003kt}, have indicated  that the observed states are most likely members of the ($1^+,2^+$) doublet. This indicates that  the positive parity doublet of $B_s$ states  ($0^+,1^+$)  is still unobserved. 
Above mentioned studies also suggest that the ($0^+,1^+$)  doublet  of $B_s$ states might  have masses below the $BK$ and $B^*K$ thresholds.
However, some relativistic quark models analysis   \cite{Ebert:2009ua,DiPierro:2001uu,Sun:2014wea} suggested  that masses of the ($0^+,1^+$)  doublet $B_s$ states  should be above the $BK$ and $B^*K$ thresholds. This reminds strongly of the 
  "history" of establishing charm meson spectrum 
in which detection of positive  parity states bellow $DK$ threshold was not predicted by the  quark models. After the observation of $D_{s 0}^*(2317)$ and  $D_{s 1}^*(2460)$ charmed mesons,  we faced  a long-lasting dilemma on the structure of $D_s$ ($0^+,1^+$). The issue is weather  the 
$D_{s 0}^*(2317)$, and  $D_{s 1}^*(2460)$ are $\bar qq$ states or more exotic  compounds \cite{Colangelo:2004vu}. 
 It was already suggested by authors of \cite{Colangelo:2004vu}, that study of strong and radiative decay modes of positive parity $D_s$ states might help in differentiating between these scenarios.  Since,  both systems of  positive parity states $D_s$ and $B_s$  are rather similar, 
 the systematic analyses of strong and radiative decay dynamics of $B_s$ mesons would help in clarifying their structure. 
In our study we rely on the  results of lattice calculations presented in Ref.  \cite{Lang:2015hza}. These authors determined the spectrum of $B_s$ $1P$ states  and they found that masses of $B_{s1}(5830)^0$  and $B^*_{s2}(5840)^0$  agree very well with experimental results  \cite{Abazov:2007af,Aaltonen:2007ah,Aaij:2012uva,Aaltonen:2013atp}. 
They predicted also existence of the spin zero positive parity state ($J^P= 0^+$) with the mass  $m_{B_{s0}} =5.711(13)(19)$ GeV and the state $J^P= 1^+$ with the mass $m_{B_{s0}} =5.750(17)(19)$ GeV.  Both states have masses below $BK$ and $B^*K$ threshold. This immediately indicates that both these states can decay strongly if isospin is violated. 
Motivated by the result of  lattice calculation and relying on our findings in the appropriate charm sector \cite{Fajfer:2015zma}, we determine partial 
decay widths of both  meson states to the final state containing one or two pions:  $B_s(0^+) \to B_s^0 \pi^0$, $B_s(1^+) \to B_s^{*0} \pi^0$ and $B_s(1^+) \to B_s^0 \pi \pi$.

Studies of these decays  were  performed already by \cite{Guo:2006rp,Guo:2006fu,Bardeen:2003kt,Lu:2006ry,
Faessler:2008vc}. Authors of \cite{Faessler:2008vc} assumed that the positive parity $0^+$ and $1^+$ $B_s$ states have a structure of $BK$ molecules, accounting for  the similarity with $D_s$, and they suggest  that they are rather narrow states with partial decay widths  about $50-60\,$ keV. On the other hand, authors of \cite{Guo:2006rp,Guo:2006fu,Bardeen:2003kt,Lu:2006ry} used different approach based on the assumption that the decays proceed  trough the channels $B_s(0^+) \to B_s^0 \eta \to B_s^0 \pi^0$ and $B_s(1^+) \to B_s^{*0} \eta \to B_s^{*0} \pi^0$  with the help of $\eta-\pi$ mixing and predicted the partial decay widths in a range of $10-40$ GeV. 
As already  discussed  in  \cite{Fajfer:2015zma,Stewart:1998ke,Fajfer:2006hi,Becirevic:2004uv}   chiral loop corrections play an important role in strong decays of $D_s$ positive parity states and their contribution to the strong decay modes can be as large as the effect of $\eta-\pi$ mixing. 
Since $\pi$ and $\pi \pi$ in the final state of these decays are having very small momenta, both decay modes are ideal to use heavy meson chiral perturbation theory (HM$\chi$PT).

In this paper, we determine  the  isospin violating decay amplitudes of positive parity $B_s$ mesons, members of the   ($0^+,1^+$) doublet,  using HM$\chi$PT.   
For two-body decays, there is a tree-level contribution to decay amplitude arising from the $\eta -\pi$ mixing and loop contribution which is the divergent. 
The divergent loop contribution requires the regularisation by the counter-terms.  On the other hand, in the isospin violating two body decays of $D_{s0}(2317)$ and $D_{s1}(2460)$ mesons,  chiral loops contribute  significantly  \cite{Fajfer:2015zma}. This was indicated already in Ref. \cite{Cleven:2014oka} within different framework in which only part of the loop contributions are included in the decay amplitudes of $D_{s0}(2317)$ and $D_{s1}(2460)$. As we pointed out in \cite{Fajfer:2015zma}, the isospin violating three body decay amplitude  can arise at the loop level only within HM$\chi$PT. These loop contributions  are then finite. 
In the case of charm decays,  the ratio of the decay widths for $D_{s1}(2460)^+\to D_s^{*+}\pi^0$ and 
$D_{s1}(2460)^+\to D_s^+ \pi^+ \pi^-$  is known experimentally. From this ratio we were able to constrain the finite size of the counter-terms necessary to regularise two body decay amplitude $D_{s1}(2460)^+ \to D_s^{*+} \pi^0$.
The heavy quark symmetry  implies the same size of counter-term contributions for $B_s$ system as in the case of charm mesons. 
Therefore, by adopting  the  result of lattice calculation  that $B_s$ mesons, part of the ($0^+,1^+$) doublet,  have masses  bellow $BK$ and $BK^*$,  we are able to predict their partial decay widths. 

The basic HM$\chi$PT formalism is introduced in Section II. In Section III we  calculate decay widths of the two body strong decays of positive parity $B_s$  doublet ($0^+$,$1^+$). In Section IV, the calculation  of the three body  decay width $B_s(1^+) \to B_s^0 \pi \pi$ decay mode will be presented,  while a short conclusion will be given in Section V.

\section{Framework}
\label{framework}

In our analysis we rely on  HM$\chi$PT  (see e.g.\cite{Burdman:1992gh,Wise:1992hn}).   This approach combines the heavy quark effective theory with the chiral perturbation theory and can be used to describe decays of mesons that are composed of one light and one heavy quark. 
The  chiral perturbation theory works very well in the case where pseudoscalar mesons have  low momenta.
In the heavy meson limit, heavy mesons, pseudoscalar and vector,  as well
as scalar and axial, become degenerate.
The negative parity states are described by the field $H$, while the positive parity states are entering in the  field $S$:
\begin{equation}
H=\frac{1}{2}(1+v \cdot \gamma)[P^*_\mu \gamma^\mu-P\gamma_5]\,, \qquad S=\frac{1}{2}(1+v \cdot \gamma)[D^*_\mu \gamma^\mu\gamma_5-D]\,,
\end{equation}
where $P^*_\mu$ and $P$ annihilate the vector and pseudoscalar mesons respectively, while  $D^*_\mu$ and $D$ annihilate the axial-vector and scalar mesons, respectively.
Within chiral perturbation theory, the light pseudoscalar mesons are accommodated into the octet  $\Sigma=\xi^2=e^{(2i\Pi/f)}$ with
\begin{equation}
\Pi=
\left(
\begin{matrix}
\pi^0/\sqrt{2}+\eta_8/\sqrt{6} & \pi^+ & K^+ \\
\pi^- & -\pi^0/\sqrt{2}+\eta_8/\sqrt{6} & K^0 \\
K^- & \bar K^0 & -2\eta_8/\sqrt{6}
\end{matrix} \right)
\label{eq-pi}
\end{equation}
and $f\sim 120\,$MeV at one loop level \cite{f}.
The leading order of the HM$\chi$PT Lagrangian, that describes the interaction of heavy and light mesons, can be written as
$${\cal L}=-
Tr[\bar H_a (iv\cdot {\cal D}_{ab}-\delta_{ab}\Delta_H)H_b]+gTr[\bar H_bH_a\gamma \cdot {\cal A}_{ab}\gamma_5]$$
\begin{equation}
+Tr[\bar S_a (iv\cdot {\cal D}_{ab})-\delta_{ab}\Delta_S)S_b]+\tilde g Tr[\bar S_b S_a\gamma \cdot {\cal A}_{ab}\gamma_5]+hTr[\bar H_bS_a\gamma \cdot{\cal A}_{ab}\gamma_5]\,,
\label{eq-lagrange}
\end{equation}
where ${\cal D}_{ab}^\mu=\delta_{ab}\partial^\mu-{\cal V}_{ab}^\mu$ is a heavy meson covariant derivative, ${\cal V}_\mu =1/2(\xi^\dagger \partial_\mu \xi +\xi \partial_\mu \xi^\dagger)$ is the light meson vector current and  ${\cal A}_\mu =i/2(\xi^\dagger \partial_\mu \xi -\xi \partial_\mu \xi^\dagger)$ is the light meson axial current. A trace is taken over spin matrices and repeated light quark flavour indices.  
All terms in (\ref{eq-lagrange}) are of the order ${\cal O}(p)$  in  the chiral power counting (see e.g.\cite{Fajfer:2006hi}). Following notation of  \cite{Lang:2015hza}, $\Delta_{SH} = \Delta_S -\Delta_H=375\, {\rm GeV}$ and  in order  to maintain well behaved  chiral expansion, we consider  that this difference is of the order of pion momentum, $\Delta_{SH}\sim {\cal O}(p)$ as in \cite{Fajfer:2006hi}. 

Light mesons are described by the Lagrangian \cite{Burdman:1992gh,Wise:1992hn},  which is of the order ${\cal O}(p^2)$ in the chiral expansion
\begin{equation}
{\cal L}_0=\frac{f^2}{8}Tr[\partial_\mu \Sigma \partial^\mu \Sigma^\dagger]+\frac{f^2\lambda_0}{4}Tr[m_q^\xi \Sigma+\Sigma m_q^\xi]\,,
\label{eq-light}
\end{equation}
with the $\lambda_0={m_\pi^2}/{(m_u+m_d)}={(m^2_{K^+}-m^2_{K_0})}/{(m_u+m_d)}={(m^2_K-m^2_\pi/2)}/{m_s}\, $. From the second term in (\ref{eq-light}), we can derive the $\eta-\pi$ mixing Lagrangian \cite{mixing1,mixing2}:
\begin{equation}
{\cal L}_{\eta-\pi_0}=\frac{m_\pi^2 (m_u-m_d)}{\sqrt{3}(m_u+m_d)}\pi_0\eta\,.
\end{equation}

The scalar (pseudoscalar) and vector (axial-vector) heavy meson propagators can be written in the  form:
\begin{equation}
\frac{i}{2(k\cdot v-\Delta_i)} \qquad {\rm and} \qquad \frac{-i(g^{\mu\nu}-v^\mu v^\nu)}{2(k\cdot v-\Delta_i)}
\end{equation} 
respectively, where $\Delta_i$ in the propagator represents the residual mass of the corresponding field. Residual masses are responsible for mass splitting of heavy meson states. The difference $\Delta_{SH}$ splits the masses of positive and negative parity states. In addition, we also have a mass splitting between $B_s$ and $B$ states as well as a mass splitting between vector (axialvector) and pseudoscalar (scalar) fields. According to \cite{PDG}, the mass  splitting between $B_s$ and $B$ states is $87\,$MeV, while the  splitting between vector and pseudoscalar states is  $45\,$MeV. Since these splittings are much smaller than  $\Delta_{SH}$, they can  be safely neglected.

The coupling constants $g$, $h$ and $\tilde g$ were already discussed by several authors and determined by several methods \cite{Colangelo:1997rp}-\cite{Colangelo:2005gb}.
We will use  recent results of  the lattice QCD: $g=0.54(3)(^{+2}_{-4})$ \cite{Becirevic:2012pf}, $\tilde g=-0.122(8)(6)$ and $h=0.84(3)(2)$ \cite{Blossier:2014vea}. The lattice results will be also used for the $B_{s0}^*$ and $B_{s1}$ masses, as well as $\Delta_{SH}$ \cite{Lang:2015hza}:
$m_{B_{s0}}=5,711(13)(19)\,$GeV,  $m_{B_{s1}}=5.75(17)(19)\,$GeV and $\Delta_{SH}=375(13)(19)\,$MeV.

In order  to absorb divergences coming from  loop integrals, one needs to include counter-terms.
Following \cite{Stewart:1998ke,Fajfer:2006hi} counter-term Lagrangian can be written as:
$$
{\cal L}_{ct}=
\lambda_1[\bar H_b\bar H_a(m_q^\xi)_{ba}]+\lambda^\prime_1[\bar H_a\bar H_a(m_q^\xi)_{bb}]
-\tilde{\lambda_1}[\bar S_b\bar S_a(m_q^\xi)_{ba}]-\tilde{\lambda^\prime_1}[\bar S_a\bar S_a(m_q^\xi)_{bb}]+
$$$$
\frac{h \kappa_1^\prime\lambda_0}{(4\pi f)^2}Tr[(\bar H S \gamma_\mu{\cal A}^\mu \gamma_5)_{ab}(m_q^\xi)_{ba}]+
\frac{h \kappa_3^\prime\lambda_0}{(4\pi f)^2}Tr[(\bar H S \gamma_\mu{\cal A}^\mu \gamma_5)_{aa}(m_q^\xi)_{bb}]+
$$$$
\frac{h\kappa_5^\prime\lambda_0}{(4\pi f)^2}Tr[\bar H_a S_a \gamma_\mu{\cal A}^\mu_{bc} \gamma_5(m_q^\xi)_{cb}]+
\frac{h\kappa_9^\prime\lambda_0}{(4\pi f)^2}Tr[\bar H_c S_a \gamma_\mu{\cal A}^\mu_{bc} \gamma_5(m_q^\xi)_{ab}]+
$$
\begin{equation}
\frac{\delta_2^\prime}{(4\pi f)^2}Tr[\bar H_a S_b i v\cdot {\cal D}_{bc}\gamma_\mu{\cal A}^\mu_{ca} \gamma_5]+
\frac{\delta_3^\prime}{(4\pi f)^2}Tr[\bar H_a S_b i \gamma_\mu\cdot {\cal D}_{bc}^\mu v \cdot {\cal A}^\mu_{ca} \gamma_5]+h.c.+\hdots\,,
\label{eq-counter-terms}
\end{equation}
where $m^\xi=(\xi m_q \xi-\xi^\dagger m_q \xi^\dagger)/2$ and $D^\alpha_{bc}A^\beta_{ca}=\partial^\alpha A^\beta_{ba}+[v^\alpha A^\beta]_{ba}$. At the given scale, the finite part of $\kappa^\prime_3$ can be absorbed into the definition of $h$. Parameters $\lambda^\prime_1$ and $\tilde{\lambda^\prime_1}$ can be absorbed into the definition of heavy meson masses by phase redefinition of $H$ and $S$, while 
$\lambda_1$ and $\tilde{\lambda_1}$ split the masses of SU(3) flavor triplets of $H_a$ and $S_a$ \cite{Stewart:1998ke,Fajfer:2006hi}. Therefore, only  contributions proportional to $\kappa^\prime_1$, $\kappa^\prime_9$, $\kappa^\prime_5$, $\delta^\prime_2$ and  $\delta^\prime_3$ will be explicitly included in the amplitudes.

\section{The amplitudes and the decay widths of two body decay modes}

\begin{figure}
\begin{center}
\includegraphics[scale=0.7]{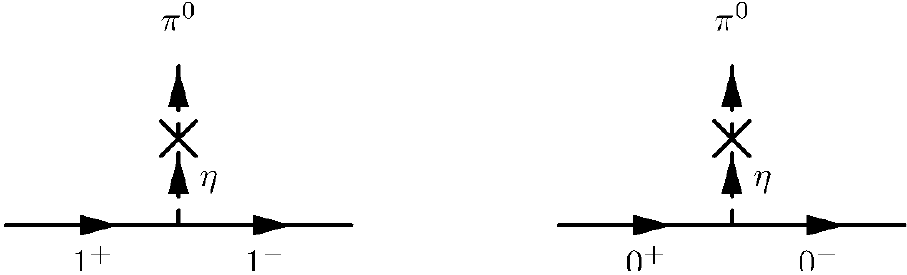}
\caption{Tree level contribution to $B^{*0}_{s0} \to B_s \pi^0$ and $B^{0}_{s1} \to B^*_s \pi^0$ decay modes.}
\label{tree}
\end{center}
\end{figure}

At the tree level, the $B^{*0}_{s0} \to B_s \pi^0$ and $B^{0}_{s1} \to B^*_s \pi^0$ decays occur 
though $\eta-\pi$ mixing as shown in Fig. $\ref{tree}$. The decay widths can be written as:
\begin{equation}
\Gamma =\frac{h^2}{2\pi f^2}|k_\pi|E_\pi^2 \delta^2_{mix}\,,
\end{equation}
where $E_{\pi}$ and $k_{\pi}$ are the energy and momenta of the outgoing pion and $\delta_{mix}$ is the 
$\eta-\pi$ mixing angle 
\begin{equation}
\delta_{mix}=\frac{1}{2\sqrt{2}}\frac{m_u-m_d}{m_s-(m_u+m_d)/2}=\frac{-1}{87\sqrt{2}}\,.
\end{equation}
This yields:
\begin{equation}
\Gamma(B^{0}_{s1} \to B^*_s \pi^0)= 16\,{\rm keV}, \qquad \qquad
\Gamma(B^{*0}_{s0} \to B_s \pi^0)=18\,{\rm keV}.
\label{eq-tree}
\end{equation}
By including  chiral loop corrections, the decay width can be rewritten as:
\begin{equation}
\Gamma=\frac{h^2}{2\pi f^2}|k_\pi|E_\pi^2 \delta^2_{mix}\left|\frac{\sqrt{Z_{w,f} Z_{w,i}}}{Z_v}\right|^2\,,
\end{equation}
where $Z_{w,f}$ and $Z_{w,i}$ denote wave function renormalization of the initial and final heavy meson states and $Z_v$ represents the vertex corrections.
\begin{figure}
\begin{center}
\includegraphics[scale=0.6]{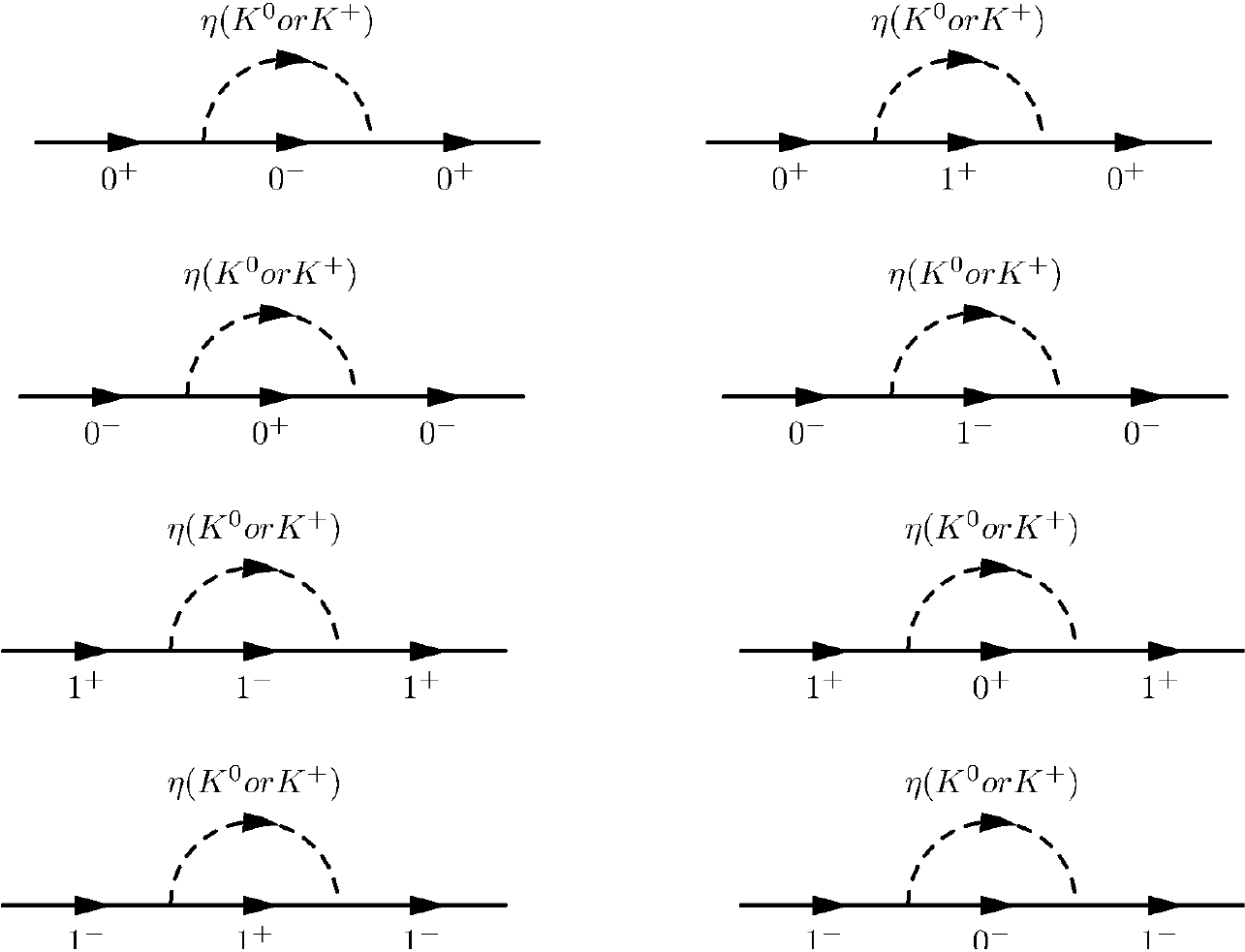}
\caption{Chiral corrections to the $B$ mesons wave functions.}
\label{sunrise}
\end{center}
\end{figure}
The wave function renormalization factor is defined as
\begin{equation}
Z_{w,j}=1-\frac{1}{2}\frac{\partial \Pi_{j} (v \cdot p)}{\partial v \cdot p} \Big |_{on\;mass\; shell}\,,
\end{equation}
where $\Pi_j(v \cdot p)$ is the meson self-energy calculated form the sunrise type diagrams in 
Fig \ref{sunrise}. 
For $Z_{w,j}$ we derive
\begin{equation}
Z_{w,j}=1-{\cal W}_j(m_{K^+})-{\cal W}_j(m_{K^0})-\frac{2}{3}{\cal W}_{j}(m_\eta)\,,
\end{equation} 
with
\begin{equation}
{\cal W}_j(m_i)=\frac{1}{16\pi^2f^2}\left(3\tilde g^2\bar B^\prime_{00}(0,m_i)
-h^2\bar B^\prime_2(-\Delta_{SH},-\Delta_{SH},m_i)\right)\,,
\end{equation}
for the positive parity mesons and
\begin{equation}
{\cal W}_j(m_i)=\frac{1}{16\pi^2f^2}\left(3 g^2 \bar B^\prime_{00}(0,m_i)
-h^2\bar B^\prime_2(\Delta_{SH},\Delta_{SH},m_i)\right)\,,
\end{equation}
for the negative parity mesons.
Here, $ \bar B^\prime_{00}$, $\bar B^\prime_2$ are Passarino-Veltman
loop integrals defined in Appendix A. 

\begin{figure}
\begin{center}
\includegraphics[scale=0.5]{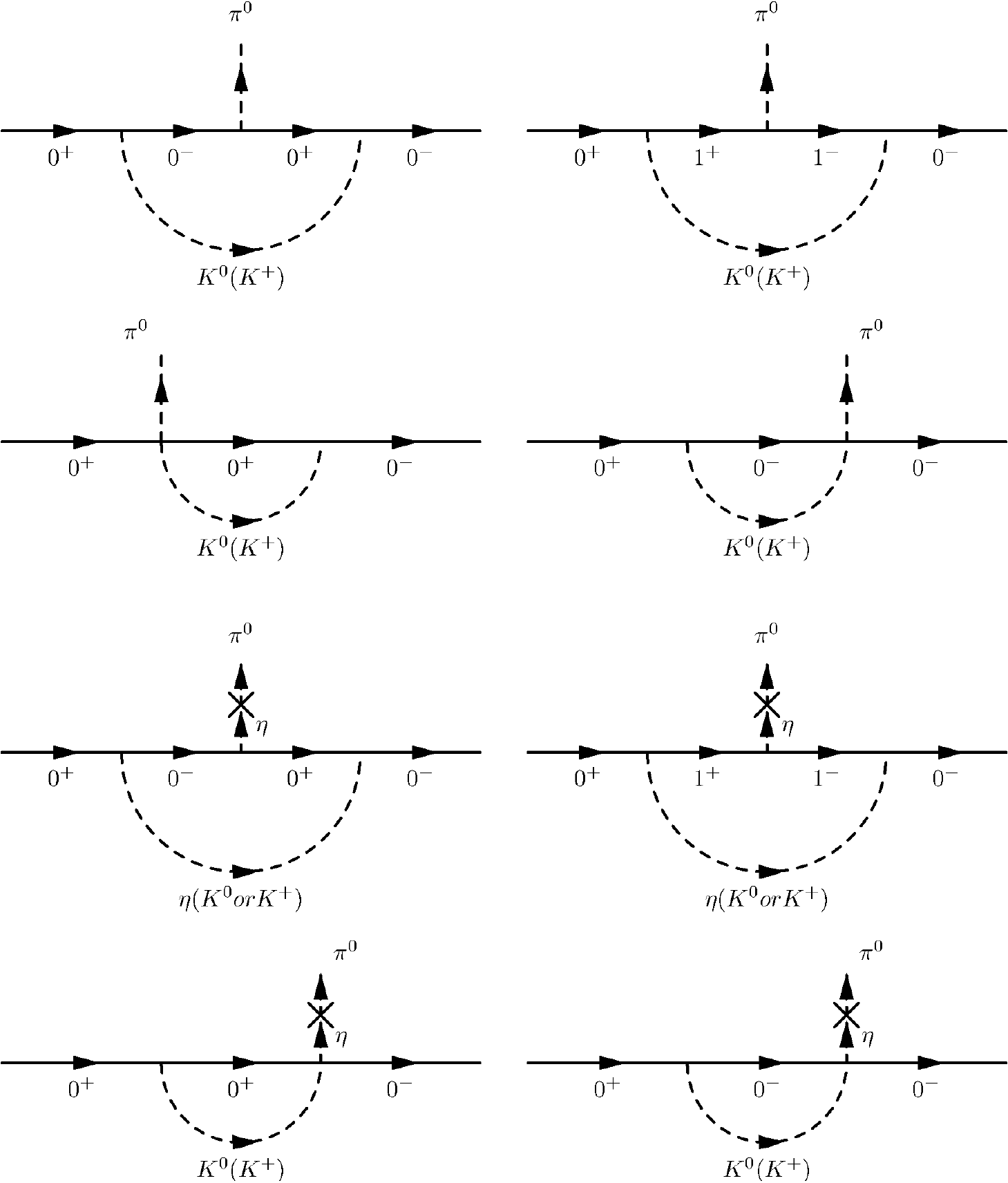}
\caption{Chiral corrections to the $B^{*0}_{s0} \to B_s \pi^0$ decay mode.}
\label{chi2}
\end{center}
\end{figure}

The vertex correction is defined as:
\begin{equation}
Z_v=1-\frac{\hat \Gamma(v\cdot p_i,v\cdot p_f,k^2)}{\hat \Gamma_{0}(v\cdot p_i,v\cdot p_f,k^2)} \Big |_{on\;mass\; shell}\,,
\end{equation}
Here $\hat \Gamma$ is the vertex amplitude calculated from the Feynman diagrams presented in Figs. \ref{chi1} and \ref{chi2}, while $\hat \Gamma_0$ is the  vertex amplitude resulting  from the  tree level Feynman diagram (see Fig. \ref{tree}):

\begin{figure}
\begin{center}
\includegraphics[scale=0.5]{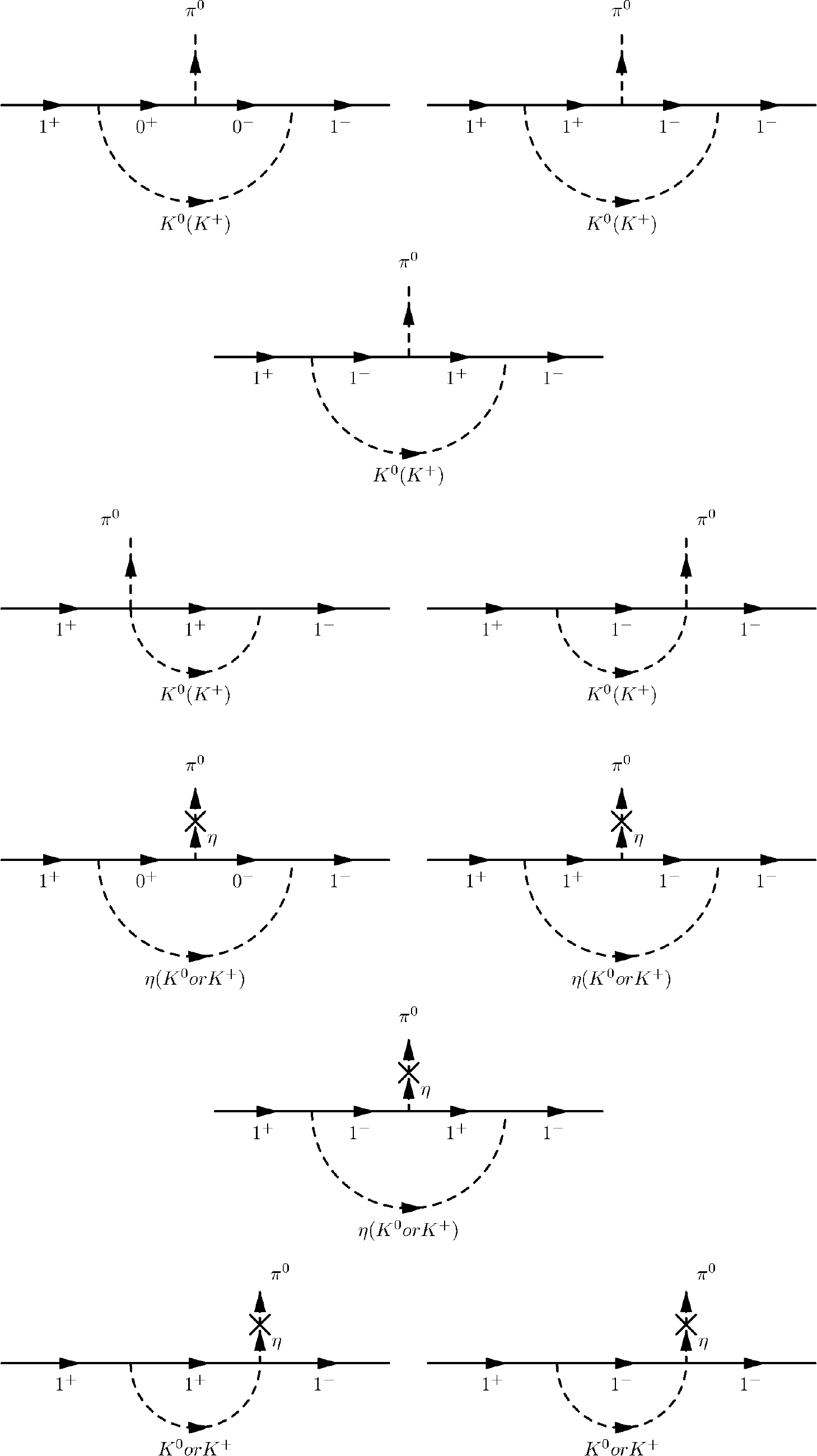}
\caption{Chiral corrections to the $B^{0}_{s1} \to B^*_s \pi^0$ decay mode.}
\label{chi1}
\end{center}
\end{figure}

\begin{equation}
Z_{v}=1-\Big(\delta^\prime_{mix}+\frac{2}{3}{\cal V}^\prime(m_\eta)-\frac{1}{2}\big({\cal V}(m_{K^+})+{\cal V}(m_{K^0})\big)+\frac{1}{\sqrt{2}\delta_{mix}}\big({\cal V}(m_{K^+})-{\cal V}(m_{K^0})\big)+{\cal V}_{ct}\Big)\,,
\end{equation}
where $\delta^\prime_{mix}=0.11$  includes corrections to the $\eta-\pi$ mixing angle beyond tree level \cite{f,Stewart:1998ke}, while ${\cal V}$ and ${\cal V}^\prime$ are

$$
{\cal V}(m_i)=\frac{1}{16\pi^2f^2}\left((\bar B_{00}(-\Delta_{SH},m_i)-\bar B_{00}(\Delta_{SH},m_i)+\bar B_{11}(-\Delta_{SH},m_i)-\bar B_{11}(\Delta_{SH},m_i)\right.
$$$$
\left.-\Delta_{SH}\bar B_{1}(-\Delta_{SH},m_i)-\Delta_{SH}\bar B_{1}(\Delta_{SH},m_i))/2\right.
$$
\begin{equation}
\left.-h^2\left(\bar B^\prime_{00}(-\Delta_{SH},\Delta_{SH},m_i)+\bar B^\prime_{11}(-\Delta_{SH},\Delta_{SH},m_i)\right)+3g\tilde g\bar B^\prime_{00}(0,0,m_i)\right)\,,
\end{equation}
\begin{equation}
{\cal V}^\prime(m_i)=\frac{1}{16\pi^2f^2}\left(
-h^2\left(\bar B^\prime_{00}(-\Delta_{SH},\Delta_{SH},m_i)+\bar B^\prime_{11}(-\Delta_{SH},\Delta_{SH},m_i)\right)+3g\tilde g\bar B^\prime_{00}(0,0,m_i)\right)\,.
\end{equation}

Note that the isospin violating nature of both decay amplitudes  manifests itself either by the proportionality of amplitude  to the mixing parameter $\delta_{mix}$, or by  the  mass difference $m_{K^0}-m_{K^+}$.
Obviously in the isospin limit, amplitudes vanish  for $\delta_{mix} \to 0$ and $ m_{K^0}=m_{K^+}$.

The  finite parts of the counter-terms are collected in the term ${\cal V}_{ct}$: 
\begin{equation}
{\cal V}_{ct}=\frac{1}{32\pi^2f^2}\left(\left(m_K^2-\frac{m_\pi^2}{2}\right)(\kappa^\prime_1+\kappa^\prime_9)+\left(m_K^2-m_\pi^2+\frac{\sqrt{2}(m^2_{K^+}-m^2_{K^0})}{\delta_{mix}}\right)\kappa^\prime_5+\frac{E_\pi}{2\lambda_0}(\delta_2^\prime+\delta_3^\prime)\right)\,.
\end{equation}

\begin{figure}
\begin{center}
\includegraphics[scale=0.5]{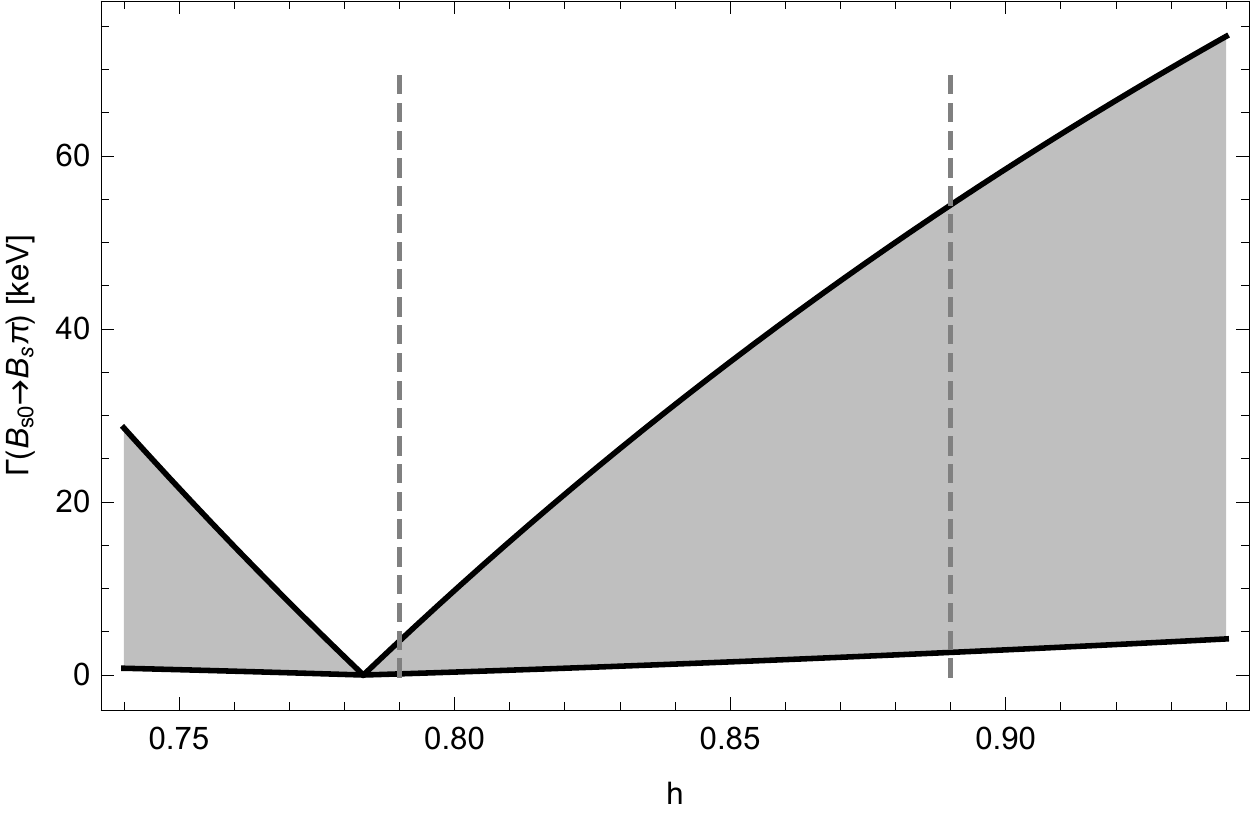}
$\qquad \qquad$
\includegraphics[scale=0.5]{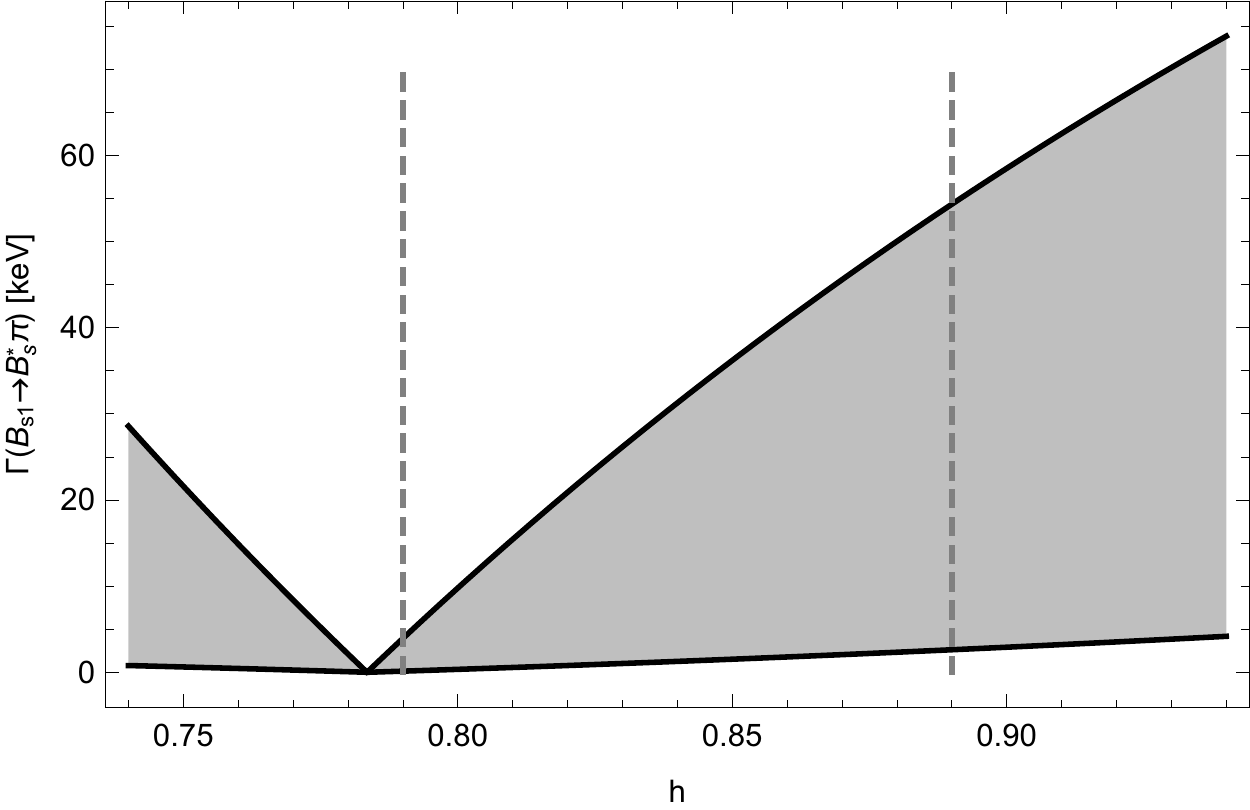}
\caption{Dependence of $\Gamma(B^0_{s1} \to B_s^0 \pi^0)$ (right) and $\Gamma(B^{*0}_{s0} \to B_s \pi^0)$ (left) on  the coupling constant $h$.}
\label{rez}
\end{center}
\end{figure}

Neglecting the terms that are multiplied by $m_\pi^2$ and $\frac{E_\pi}{2\lambda_0}$ and  by taking $m^2_{K^+}$=$m^2_{K^0}$, all counter-terms can be replaced with the linear combination
$\kappa^\prime=\kappa^\prime_1+\kappa^\prime_9+\kappa^\prime_5\,,$
yielding:
\begin{equation}
{\cal V}_{ct}=\frac{m_K^2}{32\pi^2f^2} \kappa^\prime\,.
\end{equation} 
Due to  heavy meson symmetry, the same counter-term appears also in the case of $D_s$ positive parity meson decays. In  \cite{Fajfer:2015zma} we were able to constrained the size of this counter-term using the experimentally known ratio of the decay widths of the  $D_{s1}(2460) \to D_s^*\pi$ and $D_{s1}(2460) \to D_s \pi \pi$ decay modes.  
The decay widths are also rather sensitive to the value of the coupling constant $h$ as already noticed  in \cite{Fajfer:2015zma}, for the charm meson decays. 
The wave-function renormalization factor is responsible for this behaviour. The dependence of the decay widths on the coupling constant $h$ is shown in Fig. \ref{rez}.

As seen from Fig. \ref{rez}, the decay widths are in the range of  $(0.1-55)\,$keV for the  range of coupling constant $h=-0.84(3)(2)$ as found by lattice calculation \cite{Blossier:2014vea}. 
Note that we use range of values for the counter-term $(0.1-1.2)$ as found in \cite{Fajfer:2015zma}.
For the central value  $h=0.84$,  the range is $1$ keV $\leq \Gamma(B^{0}_{s1} \to B^*_s \pi^0) \leq 30\,$keV.
The decay rates for $B^{0}_{s1} \to B^*_s \pi^0$ and $B^{*0}_{s0} \to B_s \pi^0$ are almost equal, with the small difference  due to the different masses of the final and initial $B_s$ states.

\section{The three body decays: amplitudes and decay widths}

In the case of $B^{0}_{s1}$, a three body decays  $B^{0}_{s1} \to  B^0_s \pi\pi$ are also possible.
The $B^{0}_{s1} \to  B^0_s \pi\pi$ decay width, averaged over the $B^{0}_{s1}$ polarisations, can be written as:
\begin{equation}
d\Gamma=\frac{1}{(2\pi)^3}\frac{1}{32 M_i^3}|{\cal M}|^2 dm^2_{12}dm^2_{23}\,,
\label{gama}
\end{equation}
where $M_i$ denotes the mass of  $B^{0}_{s1}$. If $p_-$ and $p_+$  are the momenta of $\pi^+$ and $\pi^-$ respectively, and $q$ is the momentum of $D_s^+$, then $dm^2_{12}=(p_++p_-)^2$ and $dm^2_{23}=(p_-+q)^2$. In the heavy quark limit $P^\mu=M_i v^\mu$,  $q^\mu=M_fv^\mu$ and 
$\epsilon \cdot v=0$, the amplitude is simplified to the following form:
$${\cal M}={\cal A}\, \epsilon \cdot (p_+-p_-)={\cal A}\, \epsilon \cdot \Delta p\,.$$

\begin{figure}
\begin{center}
\includegraphics[scale=0.5]{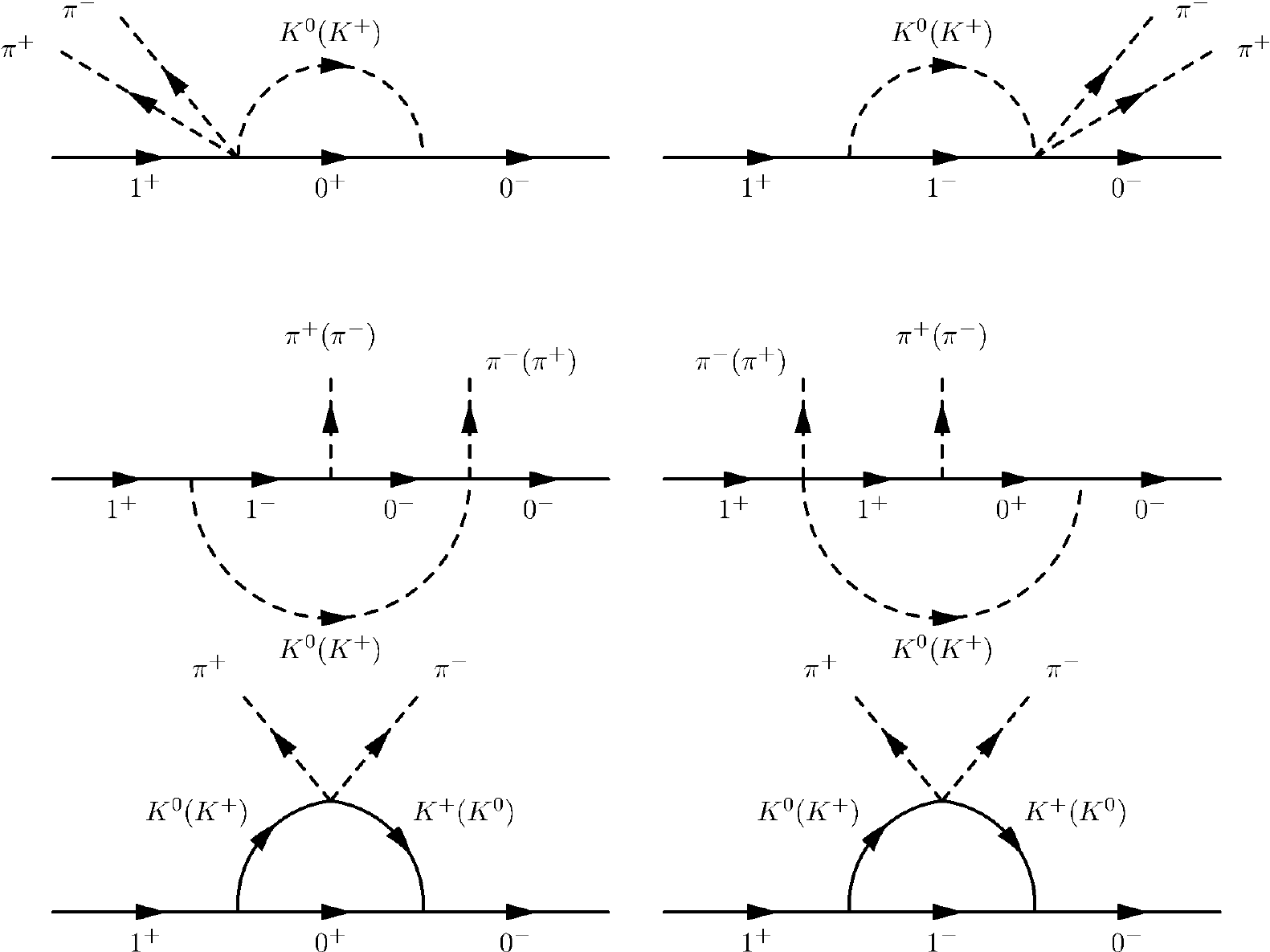}
\caption{Non-vanishing contributions  to $B_{s1}^0 \to B_s^0 \pi^+ \pi^-$ decay amplitude.} 
\label{grafi1}
\end{center}
\end{figure}

The non-vanishing Feynman diagrams that contribute to the amplitude ${\cal A}$ are presented in Fig. \ref{grafi1}. Note that all diagrams with $\eta$ meson in the loop give vanishing contribution, as it was already discussed in \cite{Fajfer:2015zma}.
The amplitude  ${\cal A}$,  can then be written as:
 
\begin{equation}
{\cal A}=\frac{h\sqrt{M_iM_f}}{16\pi^2 f^4}\left(a_1+a_2+b_1+b_2+c_1+c_2\right)\,,
\end{equation}
where parts of the amplitudes can be written as a linear combinations of the Veltman-Pasarino functions:
\begin{equation}
a_1 = \frac{g}{2}\left(\bar B_1(-\Delta_{SH}, m_{K^0})-\bar B_1(-\Delta_{SH}, m_{K^+})\right)\,,
\end{equation}
\begin{equation}
a_2 =\frac{\tilde g}{2}\left(\bar B_1(\Delta_{SH}, m_{K^0})-\bar B_1(\Delta_{SH}, m_{K^+})\right)\,,
\end{equation}
\begin{equation}
b_1 = 2g\left(\left(\bar B_2^\prime(-\Delta_{SH},-\Delta_{SH}/2, m_{K^0})
-\Delta/2\cdot \bar B_1^\prime(-\Delta_{SH},-\Delta_{SH}/2, m_{K^0})\right)\right.$$$$\left.
-\left(\bar B_2^\prime(-\Delta_{SH},-\Delta_{SH}/2, m_{K^+})
-\Delta/2\cdot \bar B_1^\prime(-\Delta_{SH},-\Delta_{SH}/2, m_{K^+})\right)
\right)\,,
\end{equation}
\begin{equation}
b_2 = 2\tilde g\left(\left(\bar B_2^\prime(\Delta_{SH}/2,\Delta_{SH}, m_{K^0})
+\Delta/2\cdot \bar B_1^\prime(\Delta_{SH}/2,\Delta_{SH}, m_{K^0})\right) \right. $$$$
-\left.\left(\bar B_2^\prime(\Delta_{SH}/2,\Delta_{SH}, m_{K^+})
+\Delta/2\cdot \bar B_1^\prime(\Delta_{SH}/2,\Delta_{SH}, m_{K^+})
\right)\right)\,,
\end{equation}
\begin{equation}
c_1=-2g\left(\left(B_{00}(m_{K^0})-\Delta_{SH}\bar C_{00}(-\Delta_{SH},m_{K^0})\right)-\left(B_{00}(m_{K^+})-\Delta_{SH}\bar C_{00}(-\Delta_{SH},m_{K^+})\right)\right)$$$$
c_2=-2\tilde g\left(B_{00}(m_{K^0})-B_{00}(m_{K^+})\right)\,.
\end{equation}

Here,  $\bar B_1$, $\bar B_2$, $B_{00}$ and $\bar C_{00}$ are the   Passarino  - Veltman loop integrals defined in Appendix A. As the $B_{s1}^0 \to B_s^0 \pi^+ \pi^-$ decay mode does not have any tree level contributions from heavy meson Lagrangian, the amplitude is expected to be finite.  
Although some of the above integrals are divergent, this divergences cancel out as expected, when we take the sum of all contributions. 
We can also notice, that the amplitude vanishes in the case of $m_{K^+}=m_{K^0}$, showing the nature of isospin violating decay mode. The obtained decay widths are: 
$$\Gamma(B_{s1}^0 \to B_s^0 \pi^+ \pi^-)=(1\pm 0.3)\times 10^{-3}\,{\rm keV}\,, \qquad \qquad
\Gamma(B_{s1}^0 \to B_s^0 \pi^0 \pi^0)=(0.7 \pm 0.2)  \times 10^{-3}\,{\rm keV}\,,$$
In the case of $B_{s1}^0 \to B_s^0 \pi^0 \pi^0$ a factor 1/2 was taken into account due to two identical mesons in the final state.

\section{Comments and discussion}

Using systematically HM$\chi$PT, we determine the decay widths of the isospin violating decay modes of positive parity $B_s$ mesons: $B_s(0^+) \to B_s^0 \pi^0$, $B_s(1^+) \to B_s^{*0} \pi^0$ and $B_s(1^+) \to B_s^0 \pi \pi$. Masses of the decaying particles and the values of the coupling constants are taken from the lattice studies. 

We find  that the decay width $\Gamma (B_s(1^+) \to B_s^0 \pi \pi) \sim 10^{-3}\,$keV.  This process occurs only at loop level  
and  the decay amplitudes are proportional to  the mass difference of $K^+$ and $K^0$. The small available phase space additionally suppresses the  decay width.
This decay might be also approached by the exchange of  the $f_0$ resonances $B_s(1^+) \to B_s^0 f_0  \to B_s^0 \pi \pi$  \cite{Bardeen:2003kt}. 
However, in the HM$\chi$PT this is  a higher order contribution   and  therefore is not considered  in our analysis. 
The approach of Ref. \cite{Bardeen:2003kt} (see Table \ref{predictions}) uses the exchange of $\sigma$ resonance in which 
there is a significant $\bar ss$ component.  However, recent lattice calculation  of \cite{Howarth:2015caa} disfavours such a content of $\sigma$.

The two body decays of $B_{s0}^{*0}$ and $B_{s1}^{*0}$  occur at  three lever trough $\eta-\pi$ mixing.
We find that the chiral loop corrections can significantly enhance or suppress decay amplitudes being  almost of the same order of magnitude as the three level contribution. 
We can only give a range of values for the  decay widths. Namely, the decay widths are very sensitive to the value of the coupling constant $h$ and change significantly if the coupling constant $h$ is varied within the error bars determined by the lattice studies \cite{Blossier:2014vea}. Also,  the counter-terms are known only within a range of values in \cite{Fajfer:2015zma}. 
In Table \ref{predictions}, we give  results of other existing  studies.  Authors of \cite{Faessler:2008vc} find higher values of decay widths in the molecular picture of positive parity $B_s$ states. In their approach, however, wave function renormalization, which in our case tends to lower the decay widths significantly, is not taken into consideration. Note also that contributions of $K^*$ loops, present in \cite{Faessler:2008vc}, are a higher order correction in HM$\chi$PT approach and therefore not included in our analysis.

It will be interesting if current experimental searches at LHCb and planned studies at Belle II would lead to discovery of both states $B_{s0}^{*0}$ and $B_{s1}^{*0}$. We hope that our study might shed more light on this issue.

\begin{table}
\begin{tabular}{lccc}
\hline \hline
method & $\Gamma(B^{*0}_{s0} \to B_s \pi^0$) [keV] $\qquad$& $\Gamma(B^{0}_{s1} \to B^*_s \pi^0)$ [keV] &$\Gamma(B^{0}_{s1} \to B_s \pi\pi$) [keV]\\ \hline \hline
molecule picture \cite{Faessler:2008vc} & 46.7 & 50.1 & \\
\hline
heavy quark and chiral symmetry \cite{Bardeen:2003kt} & 21.5 & 21.5 & $\approx$ 0.05 \\
heavy chiral unitary approach \cite{Guo:2006rp,Guo:2006fu} $\qquad$ & 7.92 & 10.36 & \\
$^3P_0$ model \cite{Lu:2006ry}  & 35 & 38 & \\
\hline
chiral loop corrections (this work) & $<55$ & $<50$ & $\approx$ 0.001 \\
\hline \hline
\end{tabular}
\caption{Predictions of the $B^{*0}_{s0} \to B_s \pi^0$ and $B^{0}_{s1} \to B^*_s \pi^0$ decay widths.}
\label{predictions}
\end{table}

.\\

{\bf ACKNOWLEDGMENTS}

The work of SF was supported in part by the Slovenian Research Agency.

\appendix

\section{Loop integrals}

By employing dimensional regularization, in the renormalization scheme with $\delta=\frac{2}{4-D}-\gamma_E+\ln 4\pi +1=0$, we have:

$$A_0(m)=\frac{(2\pi\mu)^{4-D}}{i\pi^2}\int\frac{d^Dk}{(k^2-m^2+i\epsilon)}=
m^2\left(\delta-\ln\frac{m^2}{\mu^2}\right)+{\cal O}(D-4)\,,$$
$$B_0(p,m,m)=\frac{(2\pi\mu)^{4-D}}{i\pi^2}\int\frac{d^Dk}{(k^2-m^2+i\epsilon)((k+p)^2-m^2+i\epsilon)}$$$$=\delta-\int_0^1 \ln\frac{x^2p^2-xp^2+m^2}{\mu^2}+{\cal O}(D-4)\,,$$

$$B_{00}(p,m,m)=\frac{1}{2(D-1)}[A_0(m)+(2m^2-p^2/2)B_0(p,m,m)]\,,$$
which in $D \rightarrow 4$ limit gives
$$ B_{00}(p,m,m)=\frac{1}{6}[A_0(m)+(2m^2-p^2/2)B_0(p,m,m)+2m^2-p^2/3]\,,$$
$$B_{00}(m)=B_{00}(\Delta_M v,m,m)\,.$$
Loop integrals with one heavy meson propagator are:
$$\bar B_0(\Delta,m)=\frac{(2\pi\mu)^{4-D}}{i\pi^2}\int\frac{d^Dk}{(k^2-m^2+i\epsilon)(v\cdot k-\Delta+i\epsilon)}=$$$$-2\Delta\left[\delta-\ln\frac{m^2}{\mu^2}-2F\left(\frac{m}{\Delta}\right)+1\right]+{\cal O}(D-4)\,,$$ 
with
$$F(1/x)=\Bigg\{
\begin{matrix} 
\frac{1}{x}\sqrt{x^2-1}\ln(x+\sqrt{x^2-1}+i\epsilon)\,; & |x|>1\,, \\
\frac{-1}{x}\sqrt{1-x^2}\left(\frac{\pi}{2}-\tan^{-1}\left(\frac{x}{\sqrt{1-x^2}}\right)\right)\,; & |x|\leq 1\,, 
\end{matrix}$$

$$\bar B^\mu(\Delta,m)=\frac{(2\pi\mu)^{4-D}}{i\pi^2}\int\frac{k^\mu\, d^Dk}{(k^2-m^2+i\epsilon)(v\cdot k-\Delta+i\epsilon)}=\bar B_1(\Delta,m) v^\mu\,,$$

$$\bar B_1(\Delta,m)=\frac{(2\pi\mu)^{4-D}}{i\pi^2}\int\frac{k \cdot v\, d^Dk }{(k^2-m^2+i\epsilon)(v\cdot k-\Delta+i\epsilon)}=A_0(m)+\Delta \bar B_0(\Delta,m)\,,$$

$$\bar B^{\mu\nu}(\Delta,m)=\frac{(2\pi\mu)^{4-D}}{i\pi^2}\int\frac{k^\mu k^\nu\, d^Dk}{(k^2-m^2+i\epsilon)(v\cdot k-\Delta+i\epsilon)}=\bar B_{00}(\Delta,m) g^{\mu\nu}+\bar B_{11}(\Delta,m) v^\mu v^\nu\,,$$

$$\bar B_{00}(\Delta,m)=\frac{1}{D-1}[(m^2-\Delta^2)\bar B_0(\Delta,m)-\Delta A_0(m)]\,,$$
which in $D\rightarrow 4$ gives
$$\bar B_{00}(\Delta,m)=\frac{1}{3}[(m^2-\Delta^2)\bar B_0(\Delta,m)-\Delta A_0(m)+2\Delta/3(3m^2-2\Delta^2)]\,,$$

$$\bar B_{11}(\Delta,m)=\frac{1}{D-1}[(D\Delta^2-m^2)\bar B_0(\Delta,m)+D\Delta A_0(m)]\,,$$
which in $D\rightarrow 4$ gives
$$\bar B_{11}(\Delta,m)=\frac{1}{3}[(4\Delta^2-m^2)\bar B_0(\Delta,m)+4\Delta A_0(m)-2\Delta/3(3m^2-2\Delta^2)]\,,$$
$$\bar B_{2}(\Delta,m)=\bar B_{00}(\Delta,m)+\bar B_{11}(\Delta,m)\,,$$

$$\bar B_0^\prime(\Delta_1,\Delta_2,m)=\frac{(2\pi\mu)^{4-D}}{i\pi^2}\int\frac{d^Dk}{(k^2-m^2)(v\cdot k-\Delta_1)(v\cdot k-\Delta_2)}=\frac{1}{\Delta_1-\Delta_2}[\bar B_0(\Delta_1,m)-\bar B_0(\Delta_2,m)]\,,$$

$$\bar B^{\mu\prime}(\Delta_1,\Delta_2,m)=\frac{(2\pi\mu)^{4-D}}{i\pi^2}\int\frac{k^\mu\, d^Dk}{(k^2-m^2)(v\cdot k-\Delta_1)(v\cdot k-\Delta_2)}=\bar B_1^\prime(\Delta_1,\Delta_2,m) v^\mu\,,$$

$$\bar B_1^\prime(\Delta_1,\Delta_2,m)=\frac{(2\pi\mu)^{4-D}}{i\pi^2}\int\frac{k\cdot v\, d^Dk}{(k^2-m^2)(v\cdot k-\Delta_1)(v\cdot k-\Delta_2)}=\bar B_0(\Delta_2,m)+\Delta_1 \bar B_0^\prime(\Delta_1,\Delta_2,m)\,,$$

$$\bar B_2^\prime(\Delta_1,\Delta_2,m)=\frac{(2\pi\mu)^{4-D}}{i\pi^2}\int\frac{(k\cdot v)^2\, d^Dk}{(k^2-m^2)(v\cdot k-\Delta_1)(v\cdot k-\Delta_2)}=$$$$A_0(m)+(\Delta_1+\Delta_2)\bar B_0(\Delta_2,m)+\Delta_1^2 \bar B_0^\prime(\Delta_1,\Delta_2,m)\,,$$

$$\bar B^{\mu\nu\prime}(\Delta_1,\Delta_2,m)=\frac{(2\pi\mu)^{4-D}}{i\pi^2}\int\frac{k^\mu k^\nu\, d^Dk}{(k^2-m^2)(v\cdot k-\Delta_1)(v\cdot k-\Delta_2)}=$$$$\bar B_{00}^\prime(\Delta_1,\Delta_2,m) g^{\mu\nu}+\bar B_{11}^\prime(\Delta_1,\Delta_2,m) v^\mu v^\nu\,,$$

$$\bar B_{00}^\prime(\Delta_1,\Delta_2,m)=\frac{1}{D-1}[m^2\bar B_{0}^\prime(\Delta_1,\Delta_2,m)-\Delta_1\bar B_{1}^\prime(\Delta_1,\Delta_2,m)-\bar B_{1}(\Delta_2,m)]\,,$$
which in $D\rightarrow 4$ gives
$$\frac{1}{3}[m^2\bar B_{0}^\prime(\Delta_1,\Delta_2,m)-\Delta_1\bar B_{1}^\prime(\Delta_1,\Delta_2,m)-\bar B_{1}(\Delta_2,m)+2/3(3m^ 2-2(\Delta_1^2+\Delta_2^2+\Delta_1\Delta_2))]\,,$$

$$\bar B_{11}^\prime(\Delta_1,\Delta_2,m)=\frac{1}{D-1}[-m^2\bar B_{0}^\prime(\Delta_1,\Delta_2,m)+D\Delta_1 \bar B_{1}^\prime(\Delta_1,\Delta_2,m)+D\bar B_{1}(\Delta_2,m)]\,,$$
which in $D\rightarrow 4$ gives
$$\frac{1}{3}[-m^2\bar B_{0}^\prime(\Delta_1,\Delta_2,m)+4\Delta_1 \bar B_{1}^\prime(\Delta_1,\Delta_2,m)+4\bar B_{1}(\Delta_2,m)-2/3(3m^ 2-2(\Delta_1^2+\Delta_2^2+\Delta_1\Delta_2))]\,,$$

Loop integrals with two heavy meson propagator are:
$$\bar C^\mu(p,\Delta,m_1,m_2)=\frac{(2\pi\mu)^{4-D}}{i\pi^2}\int\frac{k^\mu\, d^Dk}{(k^2-m_1^2+i\epsilon)((k-p)^2-m_2^2+i\epsilon)(v\cdot k-\Delta+i\epsilon)}=$$$$\bar C_1(p,\Delta,m_1,m_2) v^\mu\,,$$

$$\bar C_1(p,\Delta,m_1,m_2)=\frac{(2\pi\mu)^{4-D}}{i\pi^2}\int\frac{k\cdot v\, d^Dk}{(k^2-m_1^2+i\epsilon)((k-p)^2-m_2^2+i\epsilon)(v\cdot k-\Delta+i\epsilon)}=$$$$B_0(p,m_1,m_2)+\Delta\bar C_0(p,\Delta,m_1,m_2)\,,$$

$$\bar C^{\mu\nu}(p,\Delta,m_1,m_2)=\frac{(2\pi\mu)^{4-D}}{i\pi^2}\int\frac{k^\mu k^\nu\, d^Dk}{(k^2-m_1^2+i\epsilon)((k-p)^2-m_2^2+i\epsilon)(v\cdot k-\Delta+i\epsilon)}=$$$$\bar C_{00}(p,\Delta,m_1,m_2)g^{\mu\nu}+ \bar C_{11}(p,\Delta,m_1,m_2)v^\mu v^\nu\,,$$

$$\bar C_{00}(\Delta,m)=\bar C_{00}(-\Delta_M v,\Delta,m,m)=\frac{1}{D-1}[\bar B_0(-\Delta_M+\Delta,m)-(\Delta_M/2+\Delta) B_0(\Delta_m v,m,m)+$$$$(m^2-\Delta^2)\bar C_0(\Delta_M v,\Delta,m,m)]\,,$$
which in $D\rightarrow 4$ gives
$$\bar C_{00}(\Delta,m)=\bar C_{00}(-\Delta_M v,\Delta,m,m)=\frac{1}{3}[\bar B_0(-\Delta_M+\Delta,m)-(\Delta_M/2+\Delta) B_0(\Delta_m v,m,m)+$$$$(m^2-\Delta^2)\bar C_0(\Delta_M v,\Delta,m,m)-2/3(3/2\Delta_M-\Delta)]\,,$$

$$\bar C_{11}(\Delta,m)=\bar C_{11}(-\Delta_M v,\Delta,m,m)=\frac{1}{D-1}[-\bar B_0(-\Delta_M+\Delta,m)+D(\Delta_M/2+\Delta) B_0(\Delta_m v,m,m)-$$$$(m^2-D\Delta^2)\bar C_0(\Delta_M v,\Delta,m,m)]\,,$$
which in $D\rightarrow 4$ gives
$$\bar C_{11}(\Delta,m)=\bar C_{11}(-\Delta_M v,\Delta,m,m)=\frac{1}{3}[-\bar B_0(-\Delta_M+\Delta,m)+4(\Delta_M/2+\Delta) B_0(\Delta_m v,m,m)-$$$$(m^2-4\Delta^2)\bar C_0(\Delta_M v,\Delta,m,m)+2/3(3/2\Delta_M-\Delta)]\,.$$
The the calculation of the integral:
$$\bar C_0(p,\Delta,m_1,m_2)=\frac{(2\pi\mu)^{4-D}}{i\pi^2}\int\frac{d^Dk}{(k^2-m_1^2+i\epsilon)((k-p)^2-m_2^2+i\epsilon)(v\cdot k-\Delta+i\epsilon)}\,$$
is done in \cite{Zupan:2002je}.
For some calculations, we used the program FeynCalc \cite{feyncalc}.

\end{document}